\documentclass[epj,epsf,psfig]{svjour}
\usepackage{array, graphicx, subfigure}
\usepackage{graphics}
\newcommand{\gep}{G_{Ep}}
\newcommand{\gmp}{G_{Mp}}
\newcommand{\gmn}{G_{Mn}}
\newcommand{\gmpmu}{G_{Mp}/\mu_{p}}
\newcommand{\gepK}{G^{Kelly}_{Ep}}
\newcommand{\gmpK}{G^{Kelly-upd}_{Mp}}
\newcommand{\gen}{G_{En}}
\newcommand{\gmnmu}{G_{Mn}/\mu_{n}}

\newcommand{\numu}{\nu_{\mu}}
\newcommand{\muminus}{\mu^{-}}

\newcommand{\numubar}{\overline{\nu}_{\mu}}
\begin{document}
\title{Vector and Axial Nucleon Form Factors:} \subtitle{A Duality Constrained Parameterization }
\author{A. Bodek\inst{1}, S. Avvakumov\inst{1}, R. Bradford\inst{1}, and  H. Budd\inst{1}
}                     
%
%
\institute{Department of Physics and Astronomy, University of Rochester, Rochester, NY  14627-0171
}
\date{Received: date / Revised version: date}
%
\abstract{
We present new
parameterizations of vector and axial nucleon form factors.
We maintain an excellent descriptions of the
form factors at low momentum transfers, where the spatial
structure of the nucleon is important, and
use the Nachtman scaling variable $\xi$
to relate elastic and inelastic form factors
and impose quark-hadron duality constraints at high momentum
transfers where the quark structure dominates. 
We use the new vector form factors to re-extract
updated values of the axial form factor from neutrino
experiments on deuterium.
 We obtain an updated world
average value from $\numu$d
and pion electroproduction experiments of 
$M_{A}$ = $1.014 \pm 0.014~GeV/c^2$. Our parameterizations are useful 
in modeling neutrino interactions at low energies
(e.g. for neutrino oscillations experiments).  The predictions
for high momentum transfers can be tested in the 
next generation electron and neutrino scattering experiments.
\PACS{  
      {13.40.Gp}{Electromagnetic form factors}   \and
         {13.15.+g}{Neutrino interactions}   \and
            {13.85.Dz}{Elastic scattering}   \and
               {14.20.Dh}{Protons and neutrons}   \and
                    {25.30.Bf}{Elastic electron scattering}   \and
     { 25.30.Pt}{Neutrino scattering}
     } 
} 
\maketitle
\section{Introduction}
\label{intro}

The nucleon vector and axial elastic form factors have been measured
for more than 50 years in $e^- N$ and $\nu N$ scattering.
At low $Q^2$, a reasonable  description of the proton and neutron
elastic form factors is given by the dipole approximation.
The dipole approximation is a lowest-order attempt to
incorporate the non-zero size of the proton into the form
factors. The approximation assumes that the proton has
a simple exponential spatial charge distribution, $\rho(r)=\rho_0
e^{-r/r_0}$, where $r_0$ is the scale of the proton radius.
Since the form factors are related in
the non-relativistic limit to the Fourier transform of the
charge and magnetic moment distribution, the above $\rho(r)$
yields the dipole form defined by: $G_D^{V,A}(Q^2) =
{C^{V,A}}/{\left(1+\frac{Q^2}{M_{V,A}^2}\right)^2} $.
Here $C^{V,A}$= (1,$g_{A}$), $g_{A}$ = -1.267, $M_{V}^2$ = 0.71 $(GeV/c)^2$,
  and $M_{A}$ is the axial mass. 
  
Since $M_{A}$ is not equal to $M_{V}$, the 
distribution of electric and axial charge are different.
However, the magnetic moment distributions were assumed to have the
same spatial dependence as the charge distribution ({\it
i.e.}, form factor scaling). Recent measurements from Jefferson Lab 
show that the ratio of $\frac{\mu _p G_{Ep}}{G_{Mp}}$ 
falls at high $Q^2$
challenging the validity of form factor scaling \cite{jlab1} and  resulting in new updated parameterizations of
the form factors\cite{nuint05},\cite{kelly}. In this paper we present
parameterizations that simultaneously satisfy constraints at low $Q^2$ where the
spatial structure of the nucleon is important, and at high
$Q^2$ where the quark structure is important. A violation of
form-factor scaling is expected from quark-hadron
duality.
We use our
new vector form factors to re-extract updated values of the axial form factor
from a re-analysis of previous neutrino scattering data 
on deuterium and present a new parameterization for the axial form
factor within the framework of quark-hadron duality.

\begin{figure*}
 \begin{center}
\includegraphics[width=6.6in,height=4.8in]{{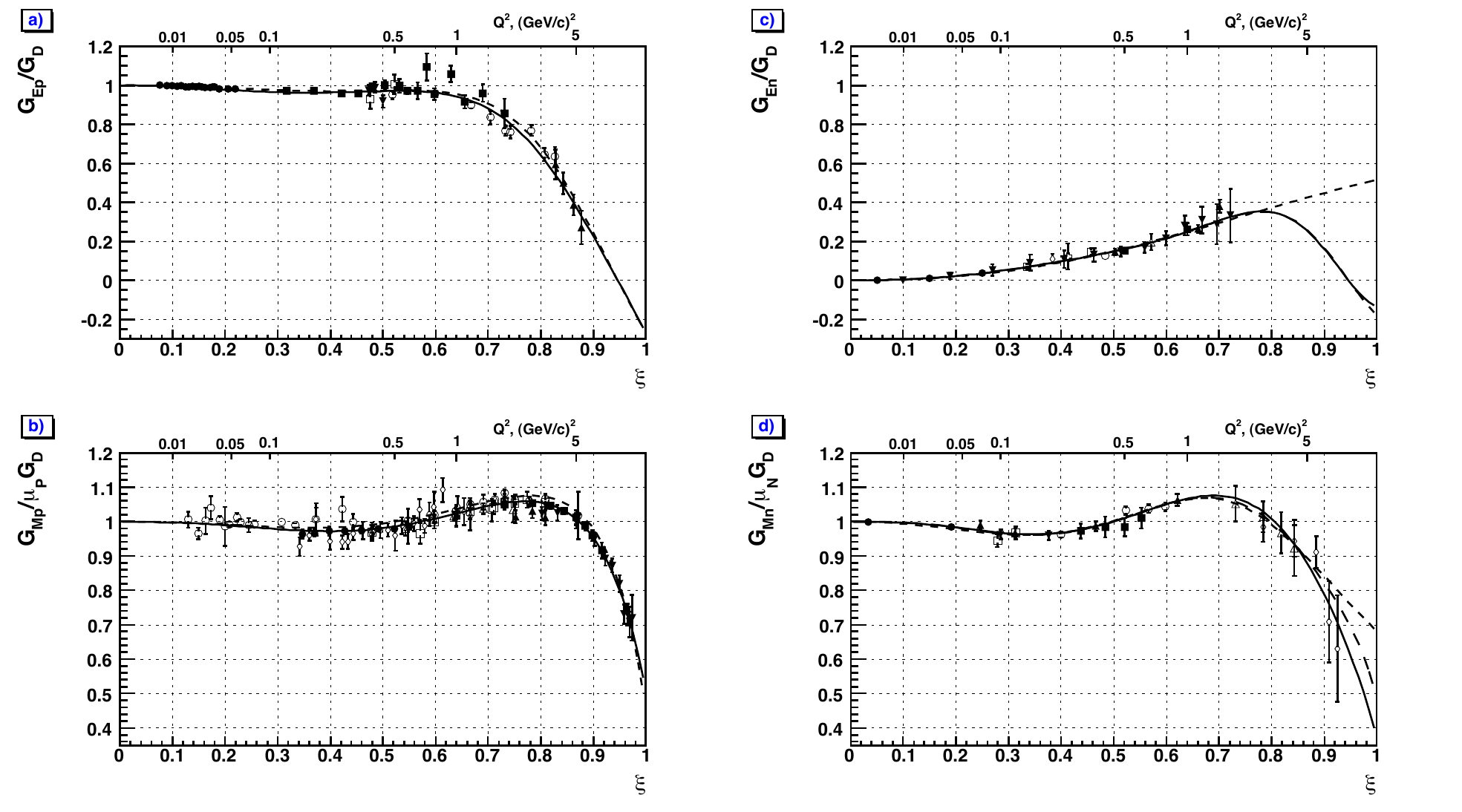}}
   \caption[Ratios to $G_D$]{Ratios of $\gep$ (a), $\gmpmu$ (b),
   $\gen$ (c) and
$\gmnmu$ (d) to $G_D$.   
The short-dashed line in each plot is
the old Kelly parameterizations (old Galster for $\gen$). 
The solid line is our new $BBBA07_{25}$
parameterization for $\frac{d}{u}=0.0$, and the long-dashed line is
 $BBBA07_{43}$ for $\frac{d}{u}=0.2$.
The values
of $\xi$ and the corresponding
values of $Q^2$ are shown on the bottom and top axis. 
}
 \end{center}
   \label{subfig1}
\end{figure*}

\section{New Parametrization}
\label{new}

The new parameterizations presented in this paper are referred to as the duality
based ``BBBA07''
parameterization. Our updated parameterizations feature the following:
(1) Improved functional form that adds an additional 
$Q^2$ dependence using the Nachtman scaling variable $\xi$
to relate elastic and inelastic
form factors.
For elastic scattering ($x=1$) 
$\xi^{p,n,N}=\frac{2}{(1+\sqrt{1+1/\tau_{p,n,N}})}$, where $\tau_{p,n,N} =
Q^2/4M_{p,n,N}^2$. Here
$M_{p,n,N}$ are the proton (0.9383 $GeV/c^2$), neutron (0.9396 $GeV/c^2$), 
and average nucleon mass (for proton,
neutron, and axial form factors, respectively).
(2) Yield the same values as
Arrington and Sick \cite{arringtonsick} for $Q^{2}< 0.64
(GeV/c)^{2}$,
while satisfying 
quark-hadron duality constraints at high-$Q^2$.

For vector form factors our fit functions are $A_N(\xi)$ (i.e. 
$A_{Ep}(\xi^{p})$, $A_{Mp}(\xi^{p})$, $A_{En}(\xi^{n})$,
$A_{Mn}(\xi^{n})$)
multiplying an updated Kelly\cite{kelly} type parameterization
 of one of the proton form factors. 
The Kelly parameterization is:
    \begin{eqnarray}
 G^{Kelly}(Q^2) = \frac{\sum_{k=0}^{m}a_k \tau^k_{p}}{1 +
\sum_{k=1}^{m+2}b_k \tau^k_{p}}, \nonumber 
\end{eqnarray}
where $a_{0}=1$ and $m=1$

\begin{table}
    \begin{center}
\begin{tabular}{l|l|l|l|l|l}
$ $ & $a_1$ & $b_1$ & $b_2$ & $b_3$ & $\chi^2/ndf$ \\
\hline
$\gepK$ & $-0.24$ & $10.98$ & $12.82$ &
$21.97$ & $
0.78$ \\
$\gmpK$ & $0.1717$ & $11.26$ & $19.32$ &
$8.33$ & $
1.03$
\end{tabular}
\caption[Kelly fit parameters]{Parameters for $G^{Kelly}_{Ep}$ and
$G^{Kelly-upd}_{Mp}$. Our parameterization employs the as-published Kelly
parameterization to $G^{Kelly}_{Ep}$ and an updated set of parameters
for $G^{Kelly-upd}_{MP}(Q^2)$ that includes the
recent BLAST\cite{crawford} results.}
%
\label{kellyparm}
\end{center}
\end{table}

In our analysis, we use all the  
datasets used by Kelly\cite{kelly}, updated to include the
recent BLAST\cite{crawford} results, to
fit $\gep$, $\gen$, $\gmpmu$, and $\gmnmu$ ($\mu_p = 2.7928$, $\mu_n = -
1.9130$).
Our parameterization employs the published Kelly
functional form to $G^{Kelly}_{Ep}$, and an updated set of parameters
for $G^{Kelly-upd}_{MP}(Q^2)$.
The 
parameters used for $G^{Kelly}_{Ep}$ and $G^{Kelly-upd}_{Mp}$ are listed in Table
\ref{kellyparm}, and  $A_N(\xi)$ is given by 
\begin{eqnarray}
 A_N (\xi) &=& \sum_{j=1}^{n} P_j (\xi) \nonumber  \\
 P_j (\xi) &=& p_j\prod_{k=1, k \ne j}^{n} \frac{\xi - \xi_k}{\xi_j -
\xi_k}. \nonumber
\end{eqnarray}
Each $P_j$ is a LaGrange polynomial in $\xi$.
  The $\xi_j$ are equidistant ``nodes'' on an interval $[0,1]$, and
$p_j$ are the
fit parameters that have an additional property $A_N (\xi_j) = p_j$.
The functional form $A_N(\xi)$ (for  $\gep$, $\gmp$, $\gen$, and $\gmn$)
is used with seven $p_j$ parameters
at $\xi_j$=0, 1/6, 1/3, 1/2, 2/3, 5/6, and 1.0.
In the fitting procedure described below, the parameters of $A_N(\xi)$ are
constrained
to give the same vector form factors as the recent low $Q^2$
fit of Arrington and Sick \cite{arringtonsick} for $Q^{2}< 0.64
(GeV/c)^{2}$ (as that analysis includes coulombs corrections which 
modify $G_{Ep}$, and two photon exchange corrections which modify
$G_{Mp}$ and $G_{Mn}$). Since the published form 
factor data do not have these corrections, this constraint is 
implemented by including additional $''fake''$ data points for $Q^{2}<
0.64 (GeV/c)^{2}$.
%
%
%
%
\begin{table*}
\begin{center}
\begin{tabular}{l|l|l|l|l|l|l|l|}
$ $ & $p_1$ & $p_2$ & $p_3$ & $p_4$ & $p_5$ &
$p_6$ & $p_7$\\
$\xi, Q^{2} $ & $0, 0$ & $0.167, 0.029$ & $0.333, 0.147$ & $0.500,
0.440$ & $0.667, 1.174,$ &
$0.833, 3.668$ & $1.0, \infty$ \\
\hline
$A_{Ep}$ &  $1.$ &$0.9927$ & $0.9898$ & $0.9975$ & $0.9812$ & $0.9340$
& $1.$  \\
$A_{Mp}$  &  $1.$  &$1.0011$ & $0.9992$ & $0.9974$ & $1.0010$ &
$1.0003$ & $1.$ \\
\hline
$A_{Ep-dipole}$ &  $1.$ &$0.9839$ & $0.9632$ & $0.9748$ & $0.9136$ &
$ 0.5447$ & $-0.2682$  \\
$A_{Mp-dipole}$  &  $1.$  &$0.9916$ & $0.9771$ & $0.9801$ & $ 1.0321$ &
$1.0429$ & $0.5084$ \\
\hline$A^{25}_{Mn}$ &  $1.$ & $0.9958$ & $0.9877$ & $1.0193$ &$1.0350$ &
$0.9164$& $0.7300$  \\
$A^{43}_{Mn}$ &  $1.$ & $0.9958$ & $0.9851$ & $1.0187$ &$1.0307 $ &
$0.9080$& $ 0.9557$  \\
\hline
$A^{25}_{En}$ & $1.$ & $1.1011$ & $1.1392$ & $1.0203$ &$1.1093$ &
$1.5429$& $0.9706$  \\
$A^{43}_{En}$ & $1.$ & $1.1019$ & $1.1387$ & $1.0234$ &$1.1046$ &
$1.5395 $& $1.2708$ \\
\hline$A^{25-dipole}_{FA}$ &$1.0000$ & $0.9207$ & $0.9795$ & $1.0480$
&$1.0516$ & $1.2874$& $ 0.7707$ \\
\end{tabular}
\caption[LaGrange Fit Parameters]{Fit parameters for $A_N(\xi)$, the
LaGrange portion of the new parameterization.  Note
$A^{25}_{Mn}$, $A^{25}_{En}$, and $A^{25}_{FA}$ are constrained
to have  $\frac{d}{u} = 0$ at $\xi = 1$,
and $A^{43}_{Mn}$, $A^{43}_{En}$,  are constrained
to have  $\frac{d}{u} = 0.2$.}
\label{lagparm}
\end{center}
\end{table*}
%
%
%
%

%

Our fits to the  form factors are:
\begin{eqnarray}
      {G_{Mp}(Q^2)}/{ \mu_{p}} &=& { A_{Mp}(\xi^{p})} \times 
      {G^{Kelly-upd}_{Mp}(Q^2)} \nonumber \\
 {G_{Ep}(Q^2)} &=& A_{Ep}(\xi^{p})\times {G^{Kelly}_{Ep}(Q^2)}
 \nonumber \\
    {G_{Mn}(Q^2)}/{\mu_{n}} &=& A^{25,43}_{Mn}(\xi^{n})
    \times  {G_{Mp}(Q^2)}/ {\mu_{p}}
 \nonumber \\   
   {G_{En}(Q^2)} &=& A^{25,43}_{En}(\xi^{n})\times {G_{Ep}(Q^2)} \times
   \left( {\frac{a\tau_{n}}{1+b\tau_{n}}} \right)
  \nonumber,
  \end{eqnarray}
 where we use 
 our updated parameters in the Kelly parameterizations.
 For $\gen$ the parameters a=1.7 and b=3.3 are the same as
 in the Galster\cite{galster}  parametrization and ensure that
 $d\gen/dQ^2$ at for $Q^{2}=0$ is in agreement with measurements.
 For convenience, we also provide
  fits for the form factors $\gep$ and $\gmpmu$ that give very
 close to the same values, but use the dipole form instead: 
 \begin{eqnarray}
     {G_{Ep}(Q^2)} &=& A_{Ep-dipole}(\xi^{p})\times {G_{D}^{V}(Q^2)} \nonumber
\\
%
  {G_{Mp}(Q^2)}/{\mu_{p}} &=&  A_{Mp-dipole}(\xi^{p})\times  {G_{D}^{V}(Q^2)} \nonumber
  \end{eqnarray}
  %
%
  
  The values  $A(\xi)$=$p_1$ at $\xi_1$=0  ($Q^2 =0$) for
   $\gmp$,  $\gep$,   $\gen$,   $\gmn$ are set to  
  to 1.0. The value  $A(\xi)$=$p_7$ at $\xi_j$=1 
  ($Q^2 \rightarrow \infty$) for $\gmp$ and $\gep$ is set to 1.0.
  
   The value  $A(\xi)$=$p_j$ at $\xi_j$=1 
for   $\gmn$ and  $\gen$ are fixed by  constraints from quark-hadron
duality.  
Quark-hadron duality implies that the ratio of neutron and
proton magnetic form factors should be the same as the ratio of the
corresponding inelastic structure functions $ \frac{F_{2n}}{F_{2p}} $
in the $\xi$=1 limit. (Here $F_2= \xi \sum_{i}e_i^2 q_i \left ( \xi \right)$)
    \begin{eqnarray}
\frac{G_{Mn}^2}{G_{Mp}^2}= \frac{F_{2n}}{F_{2p}} =
\frac{1+4\frac{d}{u}}{4+\frac{d}{u}} = \left(
\frac{\mu_{n}^{2}}{\mu_{p}^{2}}\right)
A_{Mn}^{2}(\xi=1)  \nonumber
\end{eqnarray}
 %
\begin{figure}
\includegraphics[width=3.5 in,height=3.0in]{{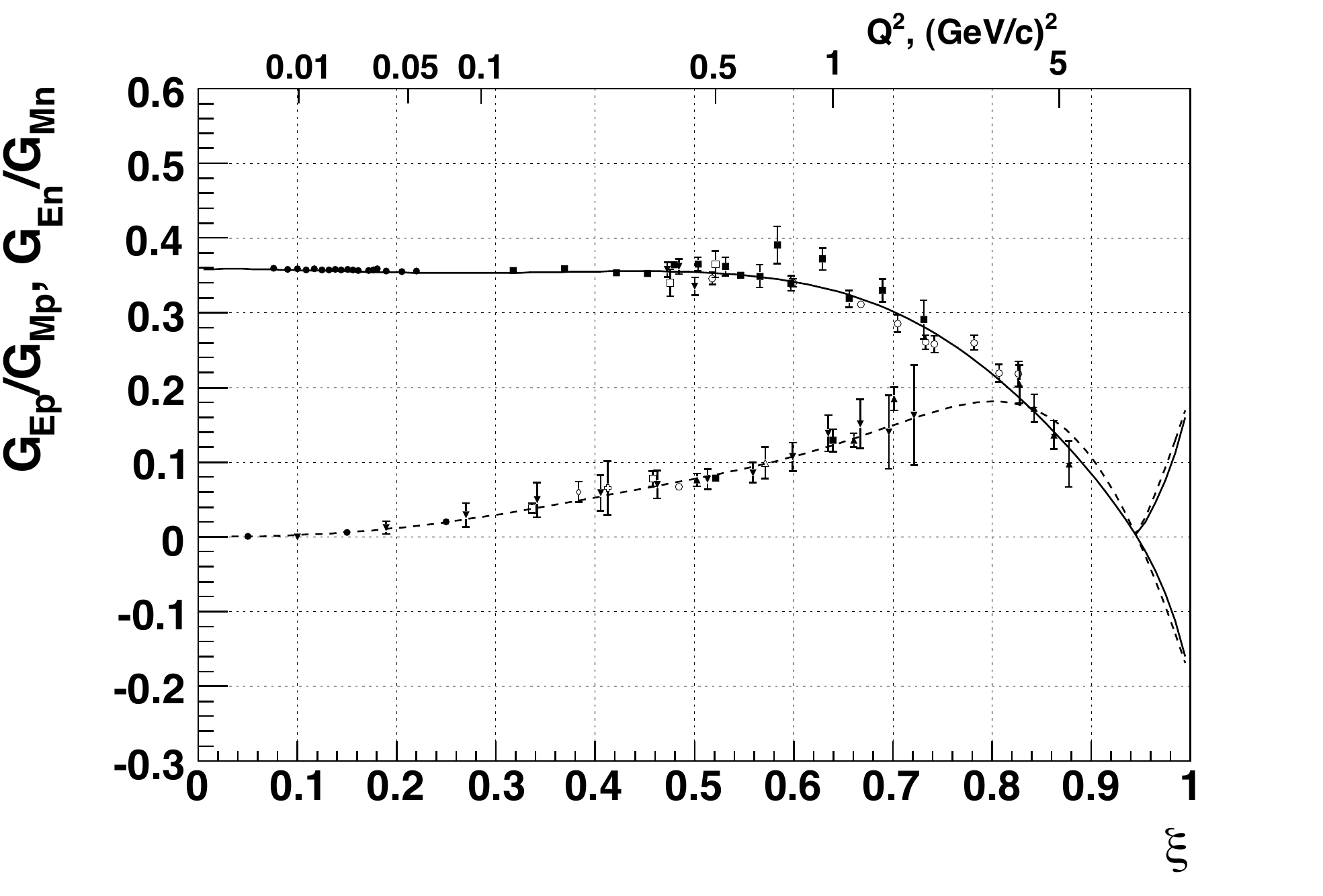}}
\caption[$G_{En}$]{The constraint used in fitting $G_{En}$
stipulates that $\gen^2/\gmn^2=\gep^2/\gmp^2$ at high $\xi$.  
The solid line is $ \frac{G_{Ep}}{|G_{Mp}|}$ and $\frac{|G_{Ep}|}{|G_{Mp}|}$, 
and the short-dashed
line is $\frac{G_{En}}{|G_{Mn}|}$  and $\frac{|G_{En}|}{|G_{Mn}|}$.
}
\label{subfig2}
\end{figure} 
%

%
\begin{table*} 
\begin{center}
\begin{tabular}{|l|l|l|l|l|l|l|l|l|
    }
    \hline 
$Experiment$ \cite{neutrinoD2} & QE & $Q^{2}$ range 
&$\overline{E}_{\nu}$ &Vector FF &
$-g_{a},M_{V}^{2}$ &   $M_A$& $\Delta M_A$  &$M_A^{updated}$	\\
$\numu$d $\rightarrow$$\muminus$ p $p_{s}$ & evnts & $GeV/c^2$ &
$GeV$ &  used \cite{ollson,galster}& used & 
(published) &  FF,RC &  $GeV/c^2$   \\
\hline 
$Mann_{73}$  & $166$ &  $.05-1.6$ & $0.7$ & $Bartl,G_{en}=0$ &1.23,$.84^{2}$ 
& 0.95 $\pm$ .12   
&   &  \\
$Barish_{77}$   & $500$ &  $.05-1.6$& $0.7$ & $Ollsn,G_{en}=0$ &1.23,$.84^{2}$
& 0.95 $\pm$ .09       & $-.026,.002$ &   \\
    $Miller_{82,77,73}$ & $1737$ &  $.05-2.5$ & $0.7$  & $Ollsn,G_{en}=0$
    &1.23,$.84^{2}$      & 1.00 $\pm$ .05      & $-.030,.002$ &
    0.972 $\pm$ .05 \\
    \hline
     $Baker_{81}$ & $1138$ &  $.06-3.0$ & $1.6$& $Ollsn,G_{en}=0$ &1.23,$.84^{2}$ 
     & 1.07 $\pm$ .06  & $-.028,.002$ &
     1.044 $\pm$ .06\\
     \hline 
 $Kitagaki_{83}$ & $362$ &  $.11-3.0$ & $20$& $Ollsn,G_{en}=0$ &1.23,$.84^{2}$   
 & 1.05$_{-.16}^{+.12}$ & $-.025,.001$
 & 1.026$_{-.16}^{+.12}$   \\ 
 \hline
 $Kitagaki_{90}$ & $2544$ &  $.10-3.0$& $1.6$ & $Ollsn,G_{en}=0$ &1.254,$.84^{2}$
 & 1.070$_{-0.045}^{+.040}$ & $-.036,.002$
 & 1.036$_{-0.045}^{+.040}$\\ 
 \hline 
  $Allasia_{90}$   & 552 & .1-3.75& $20$ &  
 $dipole,G_{en}=0$ &1.2546, $.84^{2}$ &
 $1.080 \pm .08$ 
 &$-.080,.002$ & $1.002 \pm .08$ \\
 \hline \hline
 Av.  $\numu$d \cite{bbba2007,neutrinoD2}     
 & 5780 & above & & $BBBA2007_{25}$  &1.267, .71 & $1.051 \pm .026$ 
  & $\theta_{\mu}^{-},E_{\mu},\theta,P_{p}$ &  $1.016 \pm .026$ \\  \hline
   $\pi$~$electrprd.$ \cite{pion}     &  &  &  
  & &    & & &  $1.014 \pm .016$ \\ \hline
    $\numubar$H $\rightarrow$$\muminus$ n \cite{hydrogen}    & 13 & 0-1.0&
    $1.1$ &
    $dipole,G_{en}=0$ &1.23, $.84^{2}$ &
 $0.9\pm 0.35$ &$-.070,0.01$&  $.831 \pm 0.35$ \\
 $\numubar$H $\rightarrow$$\muminus$ n \cite{hydrogen}    & 13 & 0-1.0
 & $1.1$  &
 $BBBA2007_{25}$ & 1.267, .71 &
   $\sigma_{QE}$
  & $\theta_{\mu}^{+},E_{\mu}$ &  $1.04 \pm 0.40$ \\
  \hline
   $Average~all$     &  &  &  
  & &    &  & &  $1.014 \pm .014$ \\
 \hline
\end{tabular}
\end{center}
\caption{$M_A$ $(GeV/c^2)$ values published
by  $\numu$-deuterium experiments\cite{neutrinoD2} 
and updated corrections  $\Delta M_A$ when re-extracted with updated
$BBBA2007_{25}$ form factors, and $g_{a}$=-1.267. Also 
shown is updated $M_{A}$ from $\numubar$Hydrogen 
$\rightarrow$$\muminus$ n \cite{hydrogen}. 
}
\label{MA_values-D}
\end{table*}

 We ran fits
with two different values of $\frac{d}{u}$ at the $\xi$=1 limit: 
$\frac{d}{u}$ =0 and 0.2 (corresponding
to $\frac{F_{2n}}{F_{2p}}$  = 0.25 and 0.4286). 
The fit utilizing $\frac{d}{u}=0$ is $A^{25}_{Mn}$, and
 the fit utilizing $\frac{d}{u}=0.2$ is $A^{43}_{Mn}$.  
The final parameters
 for both cases of $\frac{d}{u}$,
are given in Table \ref{lagparm} (or download computer code\cite{FF}).
The difference between these two sets is indicative of the theoretical
error of our parameterization.
Our parameterizations are within the error band of recent
theoretical fits based in dispersion relations\cite{dispersion}. Since
our fits are 
constrained
to give the same vector form factors as the recent low $Q^2$
fit of Arrington and Sick \cite{arringtonsick} for $Q^{2}< 0.64
(GeV/c)^{2}$, they are in agreement with the experimental
measurements of the proton and neutron $rms$  radii. (Note
that as discussed in reference \cite{rms}, the nucleon  $rms$
radius should be determined from fitting a polynomial of second
order to the low $Q^2$ form factors. The commonly used polynomial
of first order yields radius values which are too small).

%
%
%
The value  $A(\xi)$=$p_j$ at $\xi_j$=1 
for   $\gen$ is set by another duality-motivated
constraint.  $R$ is defined as the ratio of deep-inelastic 
longitudinal and transverse structure functions.
For inelastic scattering,
as $Q^2 \rightarrow \infty$, $R_n=R_p$. If we assume quark-hadron
duality, the same should be true for the elastic form factors
at $\xi$=1 ( $Q^2 \rightarrow \infty$) limit:
    \begin{eqnarray}
R_n\left( x=1; Q^2 \right) &=&  \frac{4M_{n}^2}{Q^2}\left( \frac{G^2
_{En}}{G^2_{Mn}}\right) \nonumber \\ 
{G^2_{En}}/{G^2_{Mn}} &=& {G^2_{Ep}}/{G^2_{Mp}} \nonumber 
\label{c2equation}
\end{eqnarray}

In order to constrain the fits to $\gen$ at high $Q^{2}$ we have
assumed that the values of  $\frac{G^2_{En}}{G^2_{Mn}}$
are the same as the measured $\frac{G^2_{Ep}}{G^2_{Mp}}$ for the
three highest $Q^{2}$ data points for $\gep$, and included these three
$''fake''$ data points in the $\gen$ fits. 
In addition, the $R_n=R_p$ condition 
yields the following constraint at $\xi = 1$:
    \begin{eqnarray}
{A^{25,43}_{En}(\xi = 1)} = P_{7} = \left( \frac{b}{a} \right)
\times \left ( \frac{1+4\frac{d}{u}}{4+\frac{d}{u}}
\right) ^{1/2}  \nonumber 
%
\end{eqnarray}
where $b/a=1.7/3.3$.  As there are two parameter sets 
$A^{25,43}_{Mn}(\xi)$, we have produced two parameter sets 
$A^{25,43}_{En}$ as shown in Table $\ref{lagparm}$.
%
  %

The new form factors $\gep$, $\gmpmu$, $\gmnmu$, and $\gen$ are plotted
in Figure 1 
as ratios to the dipole form $G_D^{V}$.

As seen in Table $\ref{lagparm}$, 
$A_N(\xi)$ is not needed for  $\gmp$ as 
it is very close to 1.0.  For $\gep$ it yields a correction
of $1\%$  at low $Q^2$ (because it is required to agree with the fits
of Arrington and Sick\cite{arringtonsick} (which include two photon exchange and Coulomb
corrections). For $\gen$ and $\gmn$  it is used to impose 
 quark-hadron duality asymptotic  constraints.
Figure \ref{subfig2} shows plots of the data
and fits to $\frac{G_{En}}{|G_{Mn}|}$ and
$\frac{G_{Ep}}{|G_{Mp}|}$ (for the  $\frac{d}{u}$ = 0 at $\xi$ = 1
case). 

 \begin{figure*}
 \begin{center}
\includegraphics[width=6.6in,height=3.0in]{{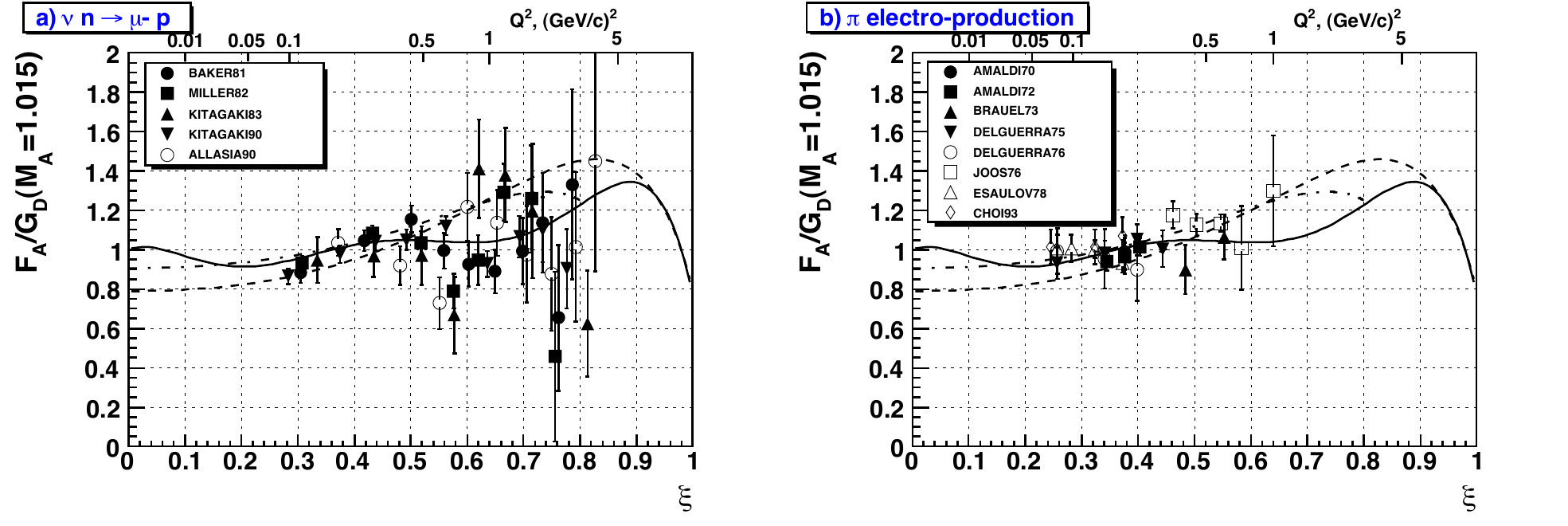}}
\caption[$F_{A}$:Axial Form factor ratios to $G_D^{A}$]{
 (a) $F_A(Q^2)$ re-extracted from neutrino-deuterium
 data divided by $G_D^{A}(Q^2)$\cite{GD}. (b)  
$F_A(Q^2)$ from pion electroproduction
  divided by $G_D^{A}(Q^2)$\cite{GD}, corrected for
  for hadronic effects\cite{pion}.
  Solid line - duality based
  fit; Short-dashed line - $F_A(Q^2)_{A2=V2}$. 
  Dashed-dot line - constituent quark model\cite{quark}.
}
 \end{center}
 %
     \label{figaxial}  
\end{figure*}

\section{Re-extraction of Axial Form Factor}
\label{axial-section}

Using our updated $BBBA2007_{25}$ form
factors and an updated
value $g_{A}$ = -1.267, we perform a complete
reanalysis of  published $\nu$ quasielastic \cite{neutrinoD2} (QE) data on
deuterium ($\numu$ n $\rightarrow$ $\muminus$p)
using the procedure described in detail
in ref. \cite{deuterium,nuint05}. 
We extract new values of $M_A$ with
updated form factors (FF) and also include
radiative corrections\cite{radcor} (RC).  
Although of lower
statistical significance, for completeness we also include all
available antineutrino data on  hydrogen targets
\cite{hydrogen}.

The average of the corrected measurements of $M_{A}$ from Table
\ref{MA_values-D}
is  $M_{A}^{deuterium}$ = $1.016 \pm 0.026$  $GeV/c^2$. 
This is in agreement the average value
of $M_{A}^{pion}$=$1.014 \pm 0.016$  $GeV/c^2$ extracted 
from pion electroproduction experiments 
after corrections for hadronic effects.\cite{pion}.
The average of the $\numu$ and electroproduction 
values is 

$M_{A}^{world-average}$=$1.014 \pm 0.014~GeV/c^{2}$.

This precise $M_{A}$ is smaller than
the recent results (for $Q^{2}$ $>$ 0.25 $(GeV/c)^2$) 
reported by MiniBoone\cite{boone} on a carbon target 
($M_{A}^{carbon}$ = 1.25 $\pm$ 0.12 $GeV/c^2$) and by the 
$K2K$\cite{k2k} collaboration on oxygen ($M_{A}^{oxygen}$ = 1.20 $\pm$ 0.12
 $GeV/c^2$). Both experiments use
updated  vector form factors. Although
the collaborations attribute the larger
 $M_{A}$ 
to nuclear effects, there are 
theoretical arguments that 
 $M_{A}$  in nuclear targets should  be smaller\cite{ma-nuclear} than 
 (or the same\cite{Tsushima_03})
 as in deuterium.
This   $M_{A}$  discrepancy is 
important for $\nu$  oscillations experiments since it
affects the normalization 
(at high energies the QE cross section
is approximately proportional to $M_{A}$) and non-linearity of the
QE cross section, which is relevant
to the extraction of  $\nu$ mass difference and mixing angle.

  For deep-inelastic scattering, the vector and axial
  parts of $F_{2}$ are equal. 
   Local quark-hadron duality 
    at large $Q^2$ implies  that
  the axial and vector components of $F_{2}^{elastic}$
  are also equal,  which yields: 

${[F_A(Q^2)_{A2=V2}]^{2}} =$

${(G_E^{V})^{2}(Q^2)+\tau_{N} 
(G_M^V(Q^2))^{2}}/{(1 + \tau_{N})},$
 %
 %
where 

$ G_E^V(Q^2)=G_{Ep}(Q^2)-G_{En}(Q^2) $,  and 
 
  $G_M^{V}(Q^2) = G_{Mp}(Q^2)-G_{Mn}(Q^2)$.

We extract values of $F_A(Q^2)$  from
the differential
 cross sections using the procedure
 of ref. \cite{deuterium}.
 The overall normalization is set by the theoretical QE
 cross section\cite{GD}.
  We  then do a duality based fit
  to 
 $F_A(Q^2)$ (including pion
 electroproduction data) of the form:
 
 $F_A (Q^2)=A^{25}_{FA} (\xi^{N}) \times G_D^{A}(Q^2).$ 
We impose the constraint  $A^{25}_{FA} (\xi_1=0) = p_1 = 1.0$. We also
constrain the fit
by requiring
that $A^{25}_{FA}(\xi^{N})$ yield  
$F_A(Q^2)=F_A(Q^2)_{A2=V2}$ by including additional $''fake''$ data
points) for $\xi > 0.9$  ($Q^{2} >7.2 (GeV/c)^2$).
 
 Figure 3(a) shows 
  $F_A(Q^2)$  extracted from neutrino-deuterium experiments 
  divided by $G_D^{A}(Q^2)$\cite{GD}.
 Figure 3(b)  shows 
 $F_A(Q^2)$  extracted from pion electroproduction
 experiments 
  divided by $G_D^{A}(Q^2)$\cite{GD}. These pion electroproduction
  values can be directly compared to the neutrino results because they
  are  multiplied  by a factor
  $F_{A}(Q^{2},M_{A}=1.014~GeV/c^{2})$/$F_{A}(Q^{2},M_{A}=1.069~GeV/c^{2})$  
  to correct  for $\Delta M_{A} = 0.055~GeV/c^{2}$ originating from 
   hadronic effects\cite{pion}.  The solid line is our duality based
 fit. The short-dashed line is  $F_A(Q^2)_{A2=V2}$. 
  The dashed-dot line
  is a constituent-quark model\cite{quark} prediction.
 
 \section{Conclusion}
\label{conclusion-section}
 
In conclusion, our new
parameterizations of vector and axial nucleon form factors
use quark-hadron duality constraints at high momentum
transfers,  and maintain a very good descriptions of the
form factors at low momentum transfers.
 Our new parameterizations
are useful in modeling $\nu$ interactions
for oscillations experiments.
Our predictions\cite{comment}
for $\gen (Q^{2}) $ and $F_{A}(Q^{2})$ 
at high $(Q^{2}$)
can be tested in future $e-N$ and $\nu$-N
 experiments.
at Jefferson Laboratory and at Fermilab (MINERvA)\cite{minerva} .


\begin{thebibliography}{}
\bibitem{comment} Local duality  may not hold for the elastic peak alone, as we
assumed, and the analysis may require using the sum of elastic and
first resonance combined.
\end{thebibliography}
\end{document}